\documentclass[10pt, a4paper]{article}

\usepackage[final]{lrec2026} 

\title{Human-Centred Evaluation of Text-to-Image Generation Models for Self-expression of Mental Distress: A Dataset Based on GPT-4o}

\name{Sui He\textsuperscript{1} and Shenbin Qian\textsuperscript{2}} 

\address{\textsuperscript{1} School of Culture and Communication, Swansea University, United Kingdom \\  \textsuperscript{2} Department of Informatics, University of Oslo, Norway \\       sui.he@swansea.ac.uk, shenbinq@ifi.uio.no \\}

\abstract{
Effective communication is central to achieving positive healthcare outcomes in mental health contexts, yet international students often face linguistic and cultural barriers that hinder their communication of mental distress. In this study, we evaluate the effectiveness of AI-generated images in supporting self-expression of mental distress. To achieve this, twenty Chinese international students studying at UK universities were invited to describe their personal experiences of mental distress. These descriptions were elaborated using GPT-4o with four persona-based prompt templates rooted in contemporary counselling practice to generate corresponding images. Participants then evaluated the helpfulness of generated images in facilitating the expression of their feelings based on their original descriptions. The resulting dataset comprises 100 textual descriptions of mental distress, 400 generated images, and corresponding human evaluation scores. Findings indicate that prompt design substantially affects perceived helpfulness, with the \textit{illustrator} persona achieving the highest ratings. This work introduces the first publicly available text-to-image evaluation dataset with human judgment scores in the mental health domain, offering valuable resources for image evaluation, reinforcement learning with human feedback, and multi-modal research on mental health communication.
 \\ \newline \Keywords{Text-to-Image Generation, Human Generated \& Evaluated Dataset, Mental Health, Self-Expression}}

\begin{document}

\maketitleabstract

\section{Introduction}

Compared to texts, visual media can evoke stronger emotional resonance and provide a more accessible channel for self-expression when communicating with others \citep{Shojaei03042025,xuetal2025}. In mental health contexts, visual and multi-modal forms of expression, such as drawings or photographs have been proven helpful for individuals to externalise complex feelings and communicate experiences that are difficult to verbalise \citep{Taylor2013-lh,Lam03102022}. Recent advances in text-to-image (T2I) generation models \citep{Ramesh2021-fl,Esser2024-nt,Labs2025-eu} extend this potential, enabling individuals to transform personal narratives into expressive and meaningful images with minimum technical expertise. 

In this broader context, international students represent a particularly vulnerable group who could benefit from such multi-modal support. Those studying in the UK, especially the vast majority with East Asian backgrounds, can face substantial challenges in expressing their feelings and/or describing their symptoms to the others, hindered by barriers imposed by their cross-cultural and inter-lingual life experiences \citep{Hyun2007-yx,Huang2020}. Research consistently shows that these barriers can lead to delayed or inadequate access to mental health support and, in severe cases, heightened the risks of isolation, academic decline, or even self-harm \citep{Clough02012019,Lazard01122016,Magnusdottir2022}. Empowering these students with multi-modal tools for self-expression can therefore help bridge communicative gaps, enabling more effective sharing of emotional experiences with peers, counsellors, or care providers who either do not share the same mother tongue or a similar lived experience.

Despite growing interest in the use of generative AI for emotional communication \citep{Hu2024,zhangetal2025}, there remains a lack of systematically constructed and publicly available datasets that combine:
	1) authentic, user-generated descriptions of mental distress,
	2) AI-generated images derived from those descriptions, and
	3) human evaluation scores reflecting perceived helpfulness of the images for self-expression.

\begin{figure}[h]
  \centering
  \includegraphics[width=0.4\textwidth]{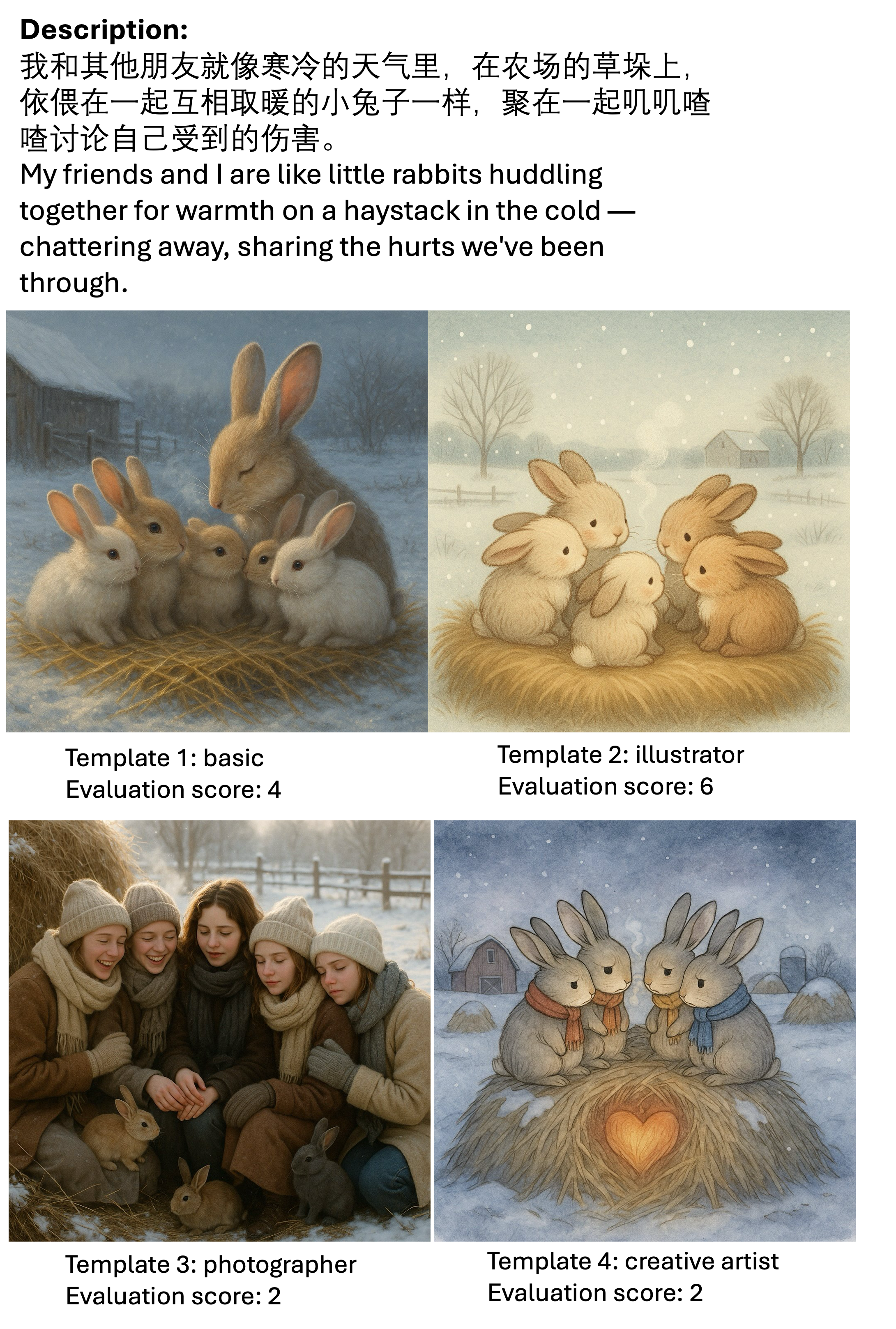}
  \caption{An example of our dataset: textual descriptions, images generated by GPT-4o and their evaluation scores. The participant described a sad experience in Chinese using a metaphor. We used the four prompt templates to ask GPT-4o to generate images, which were given categorical scores by the participant for evaluating their helpfulness.}
  \label{fig.example}
\end{figure}

To address this gap, this paper introduces a dataset of 100 descriptions of mental distress, 400 AI-generated images, and 400 categorical evaluation scores collected from 20 Chinese international students at UK universities. An example of the dataset can be seen in Figure \ref{fig.example}. The dataset provides a valuable resource for both the language resource and multi-modal communities:

\begin{itemize}
    \item Benchmarking image evaluation models in alignment with human judgments.
    \item Serving as human feedback for reinforcement learning to improve text or image generation quality.
    \item Supporting future studies on multi-modal communication and ethical AI in sensitive domains such as mental health.
\end{itemize}

Our main contributions are as follows:

\begin{itemize}
    \item \textbf{Evaluation of text-to-image generation for self-expression of mental health communication}. We conduct a systematic evaluation of GPT-4o generated images based on participants' original descriptions, including both human and automatic assessments.
    \item \textbf{Persona-based prompt templates rooted in the current counselling practices}. We design four prompt templates (i.e., basic, illustrator, photographer, creative artist) and identify which styles best align with human perceptions of helpfulness.
    \item \textbf{A novel dataset release\footnote{\url{https://anonymous.4open.science/r/T2I-Eval4MH-EA52/}}}. We release a dataset of textual descriptions, generated images, and human evaluation scores. To our knowledge, this is the first dataset in the mental health domain that provides text-to-image output with accompanying human judgments.
\end{itemize}

\section{Research Background}


When language proves insufficient, individuals often turn to visual forms of self-expression such as drawing, collage, and photography. In therapeutic contexts, art therapy and other creative practices have long been recognised for facilitating self-expression, enabling patients to externalise abstract emotions hence enhancing their communication of mental distress \citep{Shojaei03042025,xuetal2025}. Visual materials can also lower the emotional threshold for disclosure by offering a less confrontational way to articulate distress \citep{Brencio2022}. However, these traditional approaches require time, artistic skills, and access to resources, which not all help-seekers can afford. This limits their accessibility, especially for international students, who are usually young adults facing challenging situations in their host countries.

Recent advances in generative AI, particularly T2I generation models, have significantly lowered the entry level for visual creation, allowing users to produce expressive images within seconds and with minimum artistic skills. Though relatively small in number, studies to date indicate that AI-generated images can support both patients and therapists. For example, participants in \citet{zhangetal2025} reported that such images helped them articulate emotions, while \citet{Jtte2024} observed growing optimism among therapists regarding the potential of using AI-generated images in art therapy. \citet{Hu2024} further show that AI imagery can facilitate the externalisation of negative emotions in digital mental health services.

Despite the promising note that AI-generated images have shown in these studies, this topic remains underexplored, comparing to the use of other modes of communication in mental health settings. A recent review found that only 9\% of generative AI applications in healthcare involve image generation, compared with 73\% for text and 20\% for audio \citep{Xian2024-go}, not to mention the fact that empirical research on T2I generation tools for mental health is scarce to find. This highlights the novelty and urgency of exploring multi-modal AI for sensitive contexts such as self-expression of mental distress.

\section{Methodology} \label{methods}

We conducted a three-stage project consisting of 1) elicitation of mental distress descriptions via in-depth semi-structure interviews, 2) elaboration of solicited descriptions using four prompt templates and image generation, and 3) image evaluation. 

\subsection{Stage 1: Mental Distress Description}

To collect genuine descriptions from our target group, we recruited 20 Chinese international students currently studying at UK universities with self-reported experience of mental distress during their time in the UK. Semi-structured interviews were conducted online in Chinese, recorded using Zoom\footnote{\url{https://www.zoom.com/}}, and transcribed manually. Ethical approval was received from the university's ethics committee prior to participant recruitment (see Appendix~\ref{ethics} for more details). Each transcript was manually screened to select five key descriptions (100 in total) by the linguist who conducted the interviews based on their contextual knowledge and linguistic knowledge, which served as the textual inputs for image generation in Stage 2.

\subsection{Stage 2: Image Generation}

We conducted the image generation process step by step using the Chain-of-Thought prompting method \citep{wei2022}. First, we designed four prompt templates to guide GPT-4o elaborating the descriptions for visually detailed prompts to generate images: a basic prompt, and three persona-based prompts inspired by the popular concepts rooted in the studies of art therapy (\textbf{illustrator}, \textbf{photographer}, \textbf{creative artist}; see Appendix \ref{appendix}). Then, the elaborations were used in GPT-4o to generate the final images (1024×1024). Both steps were carried out using ChatGPT\footnote{\url{https://chatgpt.com/}} under the default hyperparameter setting. In total, 400 images were generated.

\subsection{Stage 3: Image Evaluation}

We conducted both human and automatic evaluation for the generated images. For human evaluation, participants in Stage 1 were invited to evaluate the images generated for each one of them through an online questionnaire. In the questionnaire, participants were asked to score the images in terms of the helpfulness of these images in facilitating their self-expression when communicating with others. We utilised Gorilla\footnote{\url{https://gorilla.sc/}}, an online platform to help pair textual descriptions with generated images for the convenience of image evaluation. Contextual information of each description was also provided on screen to help the participants recall their memory. Automatic evaluation, focusing on semantic alignment, was carried out as a complement to participants' evaluation on perceived helpfulness of these images.

\paragraph{Human Evaluation:} Each image was rated on a categorical 4-point scale (0 = \textit{not helpful}, 2 = \textit{slightly helpful}, 4 = \textit{helpful}, 6 = \textit{very helpful}), following the contrastive preference optimisation framework introduced by \citet{xuetal2024}. For each description, participants also selected the single ``best'' image from the four generated versions corresponding to the four prompt templates to reveal the effectiveness of the prompts. Along with the ``best'' image selection, a text-box was given to participants to allow them to explain the logic of scoring and their best image choices.

\paragraph{Automatic Evaluation:} We additionally computed similarity scores between the elaborations created using different prompt templates and the generated images, based on the embeddings from BLIP-2 \citep{Lietal2023} as a quantitative complement to human evaluation. We used Spearman's $\rho$ \citep{Spearman1904} and Kendall's $\tau$ \citep{kendall1938} to measure the correlation between human and automatic evaluation scores. 

\section{Results}

This section summarises our evaluation results. In Section \ref{prompt_template_re}, we demonstrate the effectiveness of our prompt templates. Section \ref{overall_re} then reports the overall findings from the human evaluation, followed by Section \ref{human_vs_automatic}, which compares the results of human and automatic evaluations.

\subsection{Effectiveness of Prompt Templates} \label{prompt_template_re}

\begin{figure}[h]
  \centering
\includegraphics[width=0.45\textwidth]{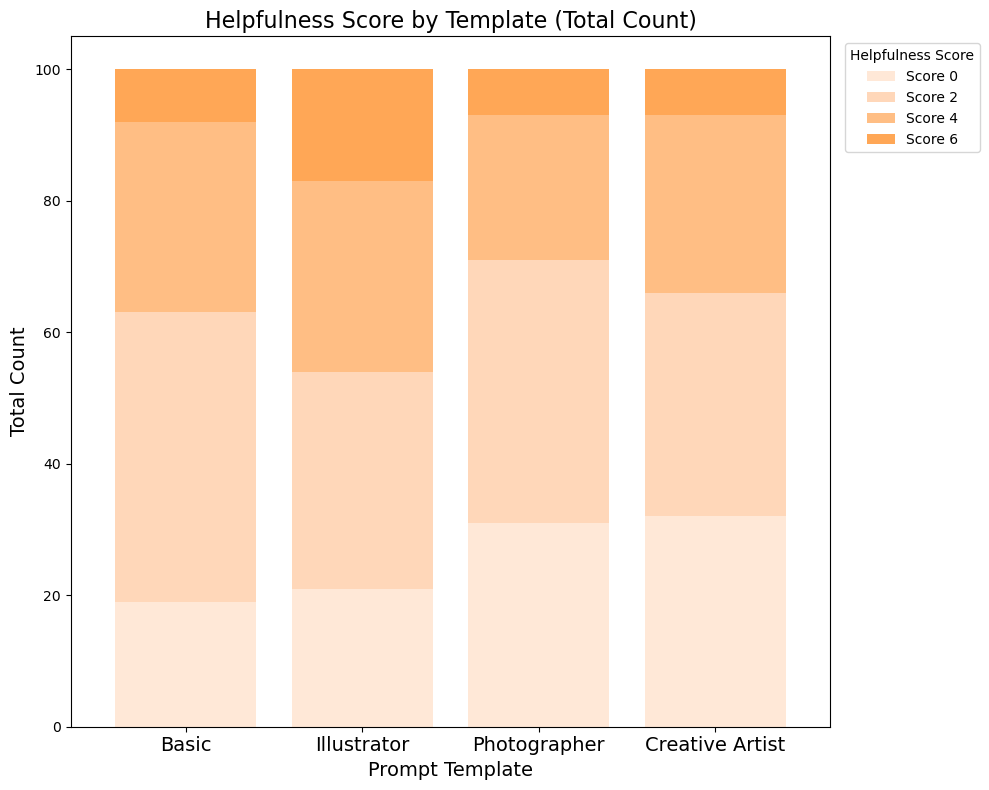}
  \caption{Degree of helpfulness of different prompt templates}
  \label{fig.prompt_prop}
\end{figure}

Figure \ref{fig.prompt_prop} shows the number of images that were rated as different degrees (categories) of helpfulness for each prompt template. It is shown that \textit{slightly helpful} and \textit{helpful} accounted for the largest percentage across all prompt templates. For the \textit{very helpful} category (6), using the `Illustrator' prompt led to the largest amount (n=17) of high image rate, whereas for the \textit{not helpful} category (0), the `Photographer' and `Creative Artist' prompts received the largest amount (n=31).

\begin{figure}[h]
  \centering
\includegraphics[width=0.45\textwidth]{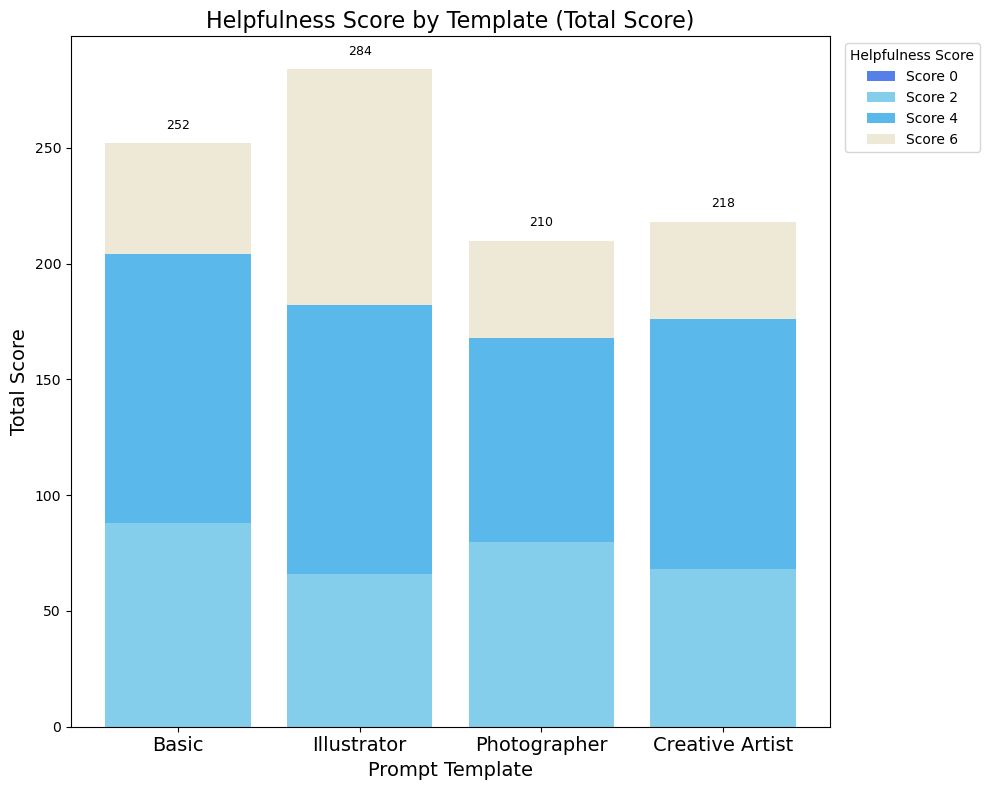}
  \caption{Total helpfulness scores for each prompt templates}
  \label{fig.prompt_num}
\end{figure}

Figure \ref{fig.prompt_num} presents the total helpfulness scores for each prompt template by summing up the score of each image generated by these prompts. Among them, the  `illustrator' prompt achieved the highest total human evaluation score (284). The basic prompt followed (252), with creative artist (218) and photographer (210) scored lower overall. Despite its lower aggregate performance, the creative artist prompt was found more effective by some participants in specific cases where the original descriptions included metaphorical or concrete imagery. For instance, a participant described their feelings about living in the UK as ``a little goldfish in a small tank''. The creative artist image received a score of 4 and was identified as their ``best'' image. 

\begin{figure}[h]
  \centering
\includegraphics[width=0.45\textwidth]{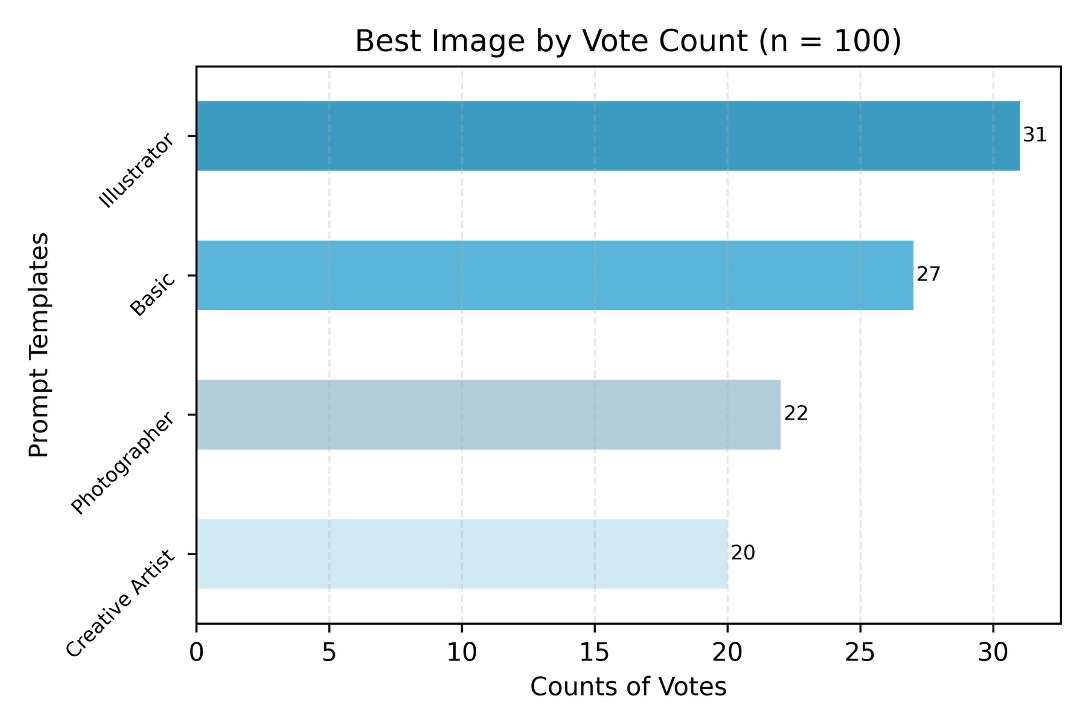}
  \caption{The number of ``best'' images selected by participants for the four prompt templates}
  \label{fig.best}
\end{figure}

\subsection{Overall Image Evaluation} \label{overall_re}

Across all 400 images, participants gave 44\% of images as \textit{slightly helpful} (2) and 27\% as \textit{helpful} (4). The mean helpfulness score was 2.4 (somewhere between \textit{slightly helpful} and \textit{helpful}), indicating a positive effect of the generated images. 

Figure \ref{fig.best} displays the number of ``best'' images selected by participants for the four prompt templates. Out of 100 ``best image'' selections, 31\% were generated by the `illustrator' prompt, 27\% by `basic', 22\% by `photographer', and 20\% by `creative artist', suggesting again that the `illustrator' prompt helps generate images rated higher by our participants. 

Analyses of the participants' image scoring logic collected from the text-box suggest that cognitive and emotional alignment of how well the image captured participants' emotional states, sense of self, or inner thoughts are the most significant consideration, as most comments related to a high score or the best image choice indicate a good match. 
 
\subsection{Human vs. Automatic Evaluation} \label{human_vs_automatic}

\begin{figure}[h]
  \centering
\includegraphics[width=0.45\textwidth]{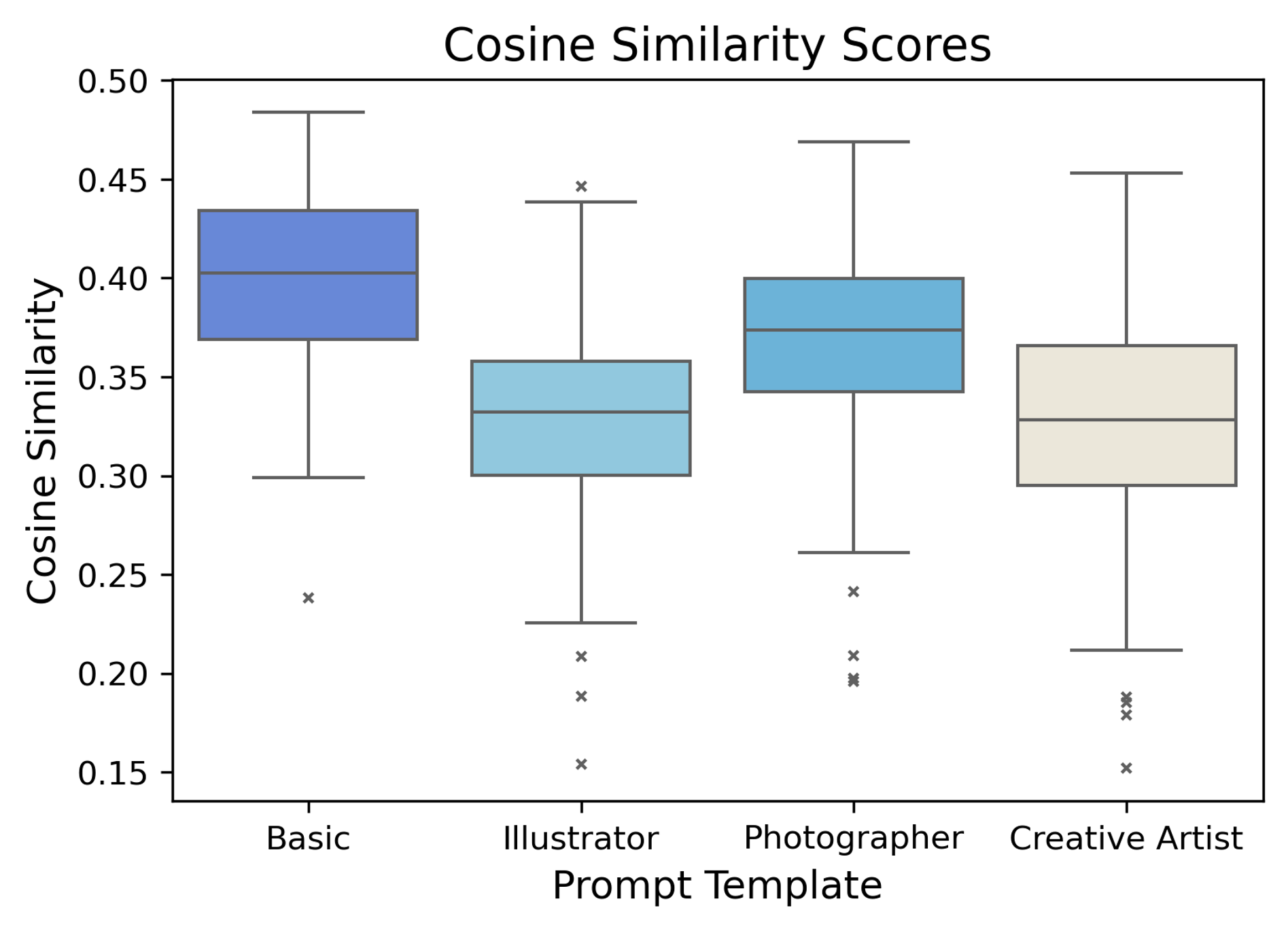}
  \caption{Similarity scores calculated using BLIP-2 embeddings for images generated based on the four prompt templates}
  \label{fig.blip2}
\end{figure}

Figure \ref{fig.blip2} shows the similarity scores calculated using BLIP-2 embeddings for images generated with the four prompt templates. We can see that the results are different from human evaluation. BLIP-2 based similarity scores prefer the `basic' prompt as images generated by this prompt have higher average score and the least outliers compared with other prompts. The Spearman and Kendall correlation scores, 0.0271 and 0.0201 also suggest that there is almost no correlation between human and automatic evaluation scores. 

This is mainly because similarity scores focus on semantic similarity between the text and the image, while human evaluation assesses to what extent the images can help express the feelings of mental distress. This difference in human and automatic evaluation highlights the value of human evaluation in contexts where emotional nuance is central to the practical function of models, and at the same time the necessity of having a novel automatic metric considering this specific application scenario. Our dataset can serve as a starting point for this endeavour.




\section{Conclusion}

This paper presents a novel dataset of 100 textual descriptions of mental distress, 400 AI-generated images, and corresponding human evaluation scores. Results indicate that although AI-generated images are by large helpful for facilitating self-expression of mental distress, prompt design can also affect their helpfulness, with the illustrator-style prompt performing best overall. This dataset provides a valuable resource for multi-modal evaluation and offers insights into how AI-generated imagery may be used to support self-expression in sensitive contexts such as mental health. For future work, we plan to conduct a large-scale human evaluation of written descriptions with comprehensive evaluation schemes in the domain of mental health through questionnaires, constructing a larger benchmark dataset for evaluating the image generation abilities of existing text-to-image generation models in producing images depicting mental distress in healthcare settings. 

\section{Acknowledgments}

We sincerely thank all the Chinese students for their participation, support, and interest in our study.

\section{Supplementary Materials}

\subsection{Appendix for Prompt Templates} \label{appendix}

\subsubsection{Basic Prompt}

\paragraph{Step 1: Elaboration}

The following task involves generating elaborations—detailed descriptions that will be used as prompts for text-to-image generation. These images are intended to support mental health communication by helping international students, whose first language is not English, express complex mental health feelings to English-speaking help-providers. 

Generate an elaboration based on the original description: [SENTENCE]. 

Keep it concise but rich in visual cues. If applicable, depict the main character as female unless stated otherwise\footnote{Most of our participants are female. For the only male participant, we changed the prompt to ask GPT-4o to depict the main character as male.}. 

\paragraph{Step 2:} Create an image (Resolution to be 1024 × 1024): [Elaboration] 

\subsubsection{Illustrator Prompt}

\paragraph{Step 1: Elaboration}

The following task involves generating elaborations—detailed descriptions that will be used as prompts for text-to-image generation. These images are intended to support mental health communication by helping international students, whose first language is not English, express complex mental health feelings to English-speaking help-providers. 

You are a professional \textbf{illustrator}. Generate an elaboration based on the original description: [SENTENCE]. 

Include the following aspects: 

– Content (what is shown) 

– Style (illustration type) 

– Colour (emotional tone or symbolism) 

– Theme (main narrative or metaphor) 

– Composition (visual layout or balance) 

Keep it concise but rich in visual cues. If applicable, depict the main character as female unless stated otherwise. 

\paragraph{Step 2:} Create an image (Resolution to be 1024 × 1024): [Elaboration] 

\subsubsection{Photographer Prompt}

\paragraph{Step 1: Elaboration}

The following task involves generating elaborations—detailed descriptions that will be used as prompts for text-to-image generation. These images are intended to support mental health communication by helping international students, whose first language is not English, express complex mental health feelings to English-speaking help-providers.  

You are a professional \textbf{photographer}. Generate an elaboration based on the original description: [SENTENCE]. 

Include the following aspects: 

– Content (what is captured) 

– Theme (main narrative or metaphor) 

– Camera View (angle, distance, or perspective) 

– Lighting (mood, source, or intensity) 

Keep it concise but rich in visual cues. If applicable, depict the main character as female unless stated otherwise. 

\paragraph{Step 2:} Create an image (Resolution to be 1024 × 1024): [Elaboration] 

\subsubsection{Creative Artist Prompt}

\paragraph{Step 1: Elaboration}

The following task involves generating elaborations—detailed descriptions that will be used as prompts for text-to-image generation. These images are intended to support mental health communication by helping international students, whose first language is not English, express complex mental health feelings to English-speaking help-providers.  

You are a \textbf{creative artist}. Generate an elaboration based on the original description: [SENTENCE]. 

Include the following aspects: 

– Content (what is depicted) 

– Style (artistic approach) 

– Theme (main narrative or metaphor) 

– Mood (overall tone or atmosphere) 

Include one unexpected or surreal element to spark creative interpretation. 

Keep it concise but rich in visual cues. If applicable, depict the main character as female unless stated otherwise. 

\paragraph{Step 2:} Create an image (Resolution to be 1024 × 1024): [Elaboration]

\subsection{Ethical considerations and limitations} \label{ethics}

\paragraph{Ethical considerations}

Ethical approval for this study was obtained from the university (approval number XXX, anonymised for review). All identifiable information was removed after the study, and personal data in the released dataset were fully anonymised. Data processing and management were conducted in accordance with GDPR guidelines.

\paragraph{Limitations} 

While this study demonstrates the potential of generative AI in supporting self-expression of mental distress, several limitations should be acknowledged. The participant group was relatively small and demographically specific, comprising 20 Chinese international students in the UK, which may limit the generalisability of findings to other cultural or linguistic groups. The use of static images captured only a single moment of expression, potentially overlooking the temporal or narrative aspects of emotional experience. In addition, the quality and resonance of generated images were influenced by both the concreteness of verbal descriptions and GPT-4o's inherent interpretive biases. Finally, technical and ethical constraints, such as the model’s safety filtering and participants’ varying familiarity with AI tools, may have shaped both image outputs and evaluation outcomes. Future work should expand the sample, include more diverse populations and modalities (e.g., video or interactive media), and explore real-time or participatory co-creation processes to enhance ecological validity.

\nocite{*}
\section{Bibliographical References}\label{sec:reference}

\bibliographystyle{lrec2026-natbib}
\bibliography{lrec2026-example}

\begin{thebibliography}{24}
\expandafter\ifx\csname natexlab\endcsname\relax\def\natexlab#1{#1}\fi

\bibitem[{Brencio(2022)}]{Brencio2022}
Francesca Brencio. 2022.
\newblock \href {https://doi.org/10.1007/978-3-030-90688-7_16} {From words to
  worlds. how metaphors and language shape mental health}.
\newblock In Shyam Wuppuluri and A.~C. Grayling, editors, \emph{Metaphors and
  Analogies in Sciences and Humanities: Words and Worlds}, pages 319--337.
  Springer International Publishing, Cham.

\bibitem[{Clough et~al.(2019)Clough, Nazareth, Day, and Casey}]{Clough02012019}
Bonnie~A. Clough, Sonia~M. Nazareth, Jamin~J. Day, and Leanne~M. Casey. 2019.
\newblock \href {https://doi.org/10.1080/03069885.2018.1459473} {A comparison
  of mental health literacy, attitudes, and help-seeking intentions among
  domestic and international tertiary students}.
\newblock \emph{British Journal of Guidance \& Counselling}, 47(1):123--135.

\bibitem[{Cogan et~al.(2024)Cogan, Liu, Chau, Kelly, Anderson, Flynn, Scott,
  Zaglis, and Corrigan}]{Cogan03072024}
Nicola~A. Cogan, X.~Liu, Y.~Chin-Van Chau, S.W. Kelly, T.~Anderson, C.~Flynn,
  L.~Scott, A.~Zaglis, and P.~Corrigan. 2024.
\newblock \href {https://doi.org/10.1080/03069885.2023.2214307} {The taboo of
  mental health problems, stigma and fear of disclosure among asian
  international students: implications for help-seeking, guidance and support}.
\newblock \emph{British Journal of Guidance \& Counselling}, 52(4):697--715.

\bibitem[{Esser et~al.(2024)Esser, Kulal, Blattmann, Entezari, M{\"u}ller,
  Saini, Levi, Lorenz, Sauer, Boesel, Podell, Dockhorn, English, Lacey,
  Goodwin, Marek, and Rombach}]{Esser2024-nt}
Patrick Esser, Sumith Kulal, Andreas Blattmann, Rahim Entezari, Jonas
  M{\"u}ller, Harry Saini, Yam Levi, Dominik Lorenz, Axel Sauer, Frederic
  Boesel, Dustin Podell, Tim Dockhorn, Zion English, Kyle Lacey, Alex Goodwin,
  Yannik Marek, and Robin Rombach. 2024.
\newblock \href {http://arxiv.org/abs/2403.03206} {Scaling rectified flow
  transformers for high-resolution image synthesis}.
\newblock \emph{arXiv preprint}.

\bibitem[{Horvat and Kis‐Rigo(2014)}]{Horvat14}
Horey D Romios~P Horvat, L and J~Kis‐Rigo. 2014.
\newblock \href {https://doi.org/10.1002/14651858.CD009405.pub2} {Cultural
  competence education for health professionals}.
\newblock \emph{Cochrane Database of Systematic Reviews}, (5).

\bibitem[{Hu et~al.(2024)Hu, Lin, Zhang, and Ji}]{Hu2024}
Chao Hu, Zhicheng Lin, Ning Zhang, and Li~Jun Ji. 2024.
\newblock \href {https://doi.org/10.3389/fpsyt.2024.1434172} {Ai-empowered
  imagery writing: integrating ai-generated imagery into digital mental health
  service}.
\newblock \emph{Frontiers in Psychiatry}, 15:01--05.

\bibitem[{Huang et~al.(2020)Huang, Kern, and Oades}]{Huang2020}
Lanxi Huang, Margaret~L. Kern, and Lindsay~G. Oades. 2020.
\newblock \href {https://doi.org/10.3390/ijerph17155538} {Strengthening
  university student wellbeing: Language and perceptions of chinese
  international students}.
\newblock \emph{International Journal of Environmental Research and Public
  Health}, 17(15).

\bibitem[{Hyun et~al.(2007)Hyun, Quinn, Madon, and Lustig}]{Hyun2007-yx}
Jenny Hyun, Brian Quinn, Temina Madon, and Steve Lustig. 2007.
\newblock Mental health need, awareness, and use of counseling services among
  international graduate students.
\newblock \emph{J Am Coll Health}, 56(2):109--118.

\bibitem[{Jütte et~al.(2024)Jütte, Wang, Steven, and Roth}]{Jtte2024}
Lennart Jütte, Ning Wang, Martin Steven, and Bernhard Roth. 2024.
\newblock \href {https://doi.org/10.3390/ai5030080} {Perspectives for
  generative ai-assisted art therapy for melanoma patients}.
\newblock \emph{AI}, 5:1648--1669.

\bibitem[{Kendall(1938)}]{kendall1938}
M.~G. Kendall. 1938.
\newblock \href {https://doi.org/10.1093/biomet/30.1-2.81} {A new measure of
  rank correlation}.
\newblock \emph{Biometrika}, 30(1-2):81--93.

\bibitem[{Labs et~al.(2025)Labs, Batifol, Blattmann, Boesel, Consul, Diagne,
  Dockhorn, English, English, Esser, Kulal, Lacey, Levi, Li, Lorenz,
  M{\"u}ller, Podell, Rombach, Saini, Sauer, and Smith}]{Labs2025-eu}
Black~Forest Labs, Stephen Batifol, Andreas Blattmann, Frederic Boesel, Saksham
  Consul, Cyril Diagne, Tim Dockhorn, Jack English, Zion English, Patrick
  Esser, Sumith Kulal, Kyle Lacey, Yam Levi, Cheng Li, Dominik Lorenz, Jonas
  M{\"u}ller, Dustin Podell, Robin Rombach, Harry Saini, Axel Sauer, and Luke
  Smith. 2025.
\newblock \href {http://arxiv.org/abs/2506.15742} {{FLUX.1} kontext: Flow
  matching for in-context image generation and editing in latent space}.
\newblock \emph{arXiv preprint}.

\bibitem[{Lam et~al.(2022)Lam, Huang, and Shen}]{Lam03102022}
Chervin Lam, Zhongwei Huang, and Liang Shen. 2022.
\newblock \href {https://doi.org/10.1080/10810730.2022.2157909} {Infographics
  and the elaboration likelihood model (elm): Differences between visual and
  textual health messages}.
\newblock \emph{Journal of Health Communication}, 27(10):737--745.
\newblock PMID: 36519844.

\bibitem[{Lazard et~al.(2016)Lazard, Bamgbade, Sontag, and
  Brown}]{Lazard01122016}
Allison~J. Lazard, Benita~A. Bamgbade, Jennah~M. Sontag, and Carolyn Brown.
  2016.
\newblock \href {https://doi.org/10.1080/10810730.2016.1245374} {Using visual
  metaphors in health messages: A strategy to increase effectiveness for mental
  illness communication}.
\newblock \emph{Journal of Health Communication}, 21(12):1260--1268.
\newblock PMID: 27869576.

\bibitem[{Li et~al.(2023)Li, Li, Savarese, and Hoi}]{Lietal2023}
Junnan Li, Dongxu Li, Silvio Savarese, and Steven Hoi. 2023.
\newblock Blip-2: bootstrapping language-image pre-training with frozen image
  encoders and large language models.
\newblock In \emph{Proceedings of the 40th International Conference on Machine
  Learning}, ICML'23. JMLR.org.

\bibitem[{Magnusdottir and Thornicroft(2022)}]{Magnusdottir2022}
Erla Magnusdottir and Graham Thornicroft. 2022.
\newblock \href {https://doi.org/10.3310/nihropenres.13268.1} {Mental health of
  chinese international students: narrative review of experiences in the uk}.
\newblock \emph{NIHR Open Research}, 52:1--18.

\bibitem[{Ramesh et~al.(2021)Ramesh, Pavlov, Goh, Gray, Voss, Radford, Chen,
  and Sutskever}]{Ramesh2021-fl}
Aditya Ramesh, Mikhail Pavlov, Gabriel Goh, Scott Gray, Chelsea Voss, Alec
  Radford, Mark Chen, and Ilya Sutskever. 2021.
\newblock \href {http://arxiv.org/abs/2102.12092} {Zero-shot text-to-image
  generation}.
\newblock \emph{arXiv preprint}.

\bibitem[{Shojaei et~al.(2025)Shojaei, Torres, and Shih}]{Shojaei03042025}
Fereshtehossadat Shojaei, John~Osorio Torres, and Patrick~C. Shih. 2025.
\newblock \href {https://doi.org/10.1080/07421656.2024.2383826} {Exploring the
  integration of technology in art therapy: Insights from interviews with art
  therapists}.
\newblock \emph{Art Therapy}, 42(2):112--118.

\bibitem[{Spearman(1904)}]{Spearman1904}
Charles Spearman. 1904.
\newblock \href {http://www.jstor.org/stable/1412159?origin=JSTOR-pdf} {The
  proof and measurement of association between two things}.
\newblock \emph{The American Journal of Psychology}, 15:72--101.

\bibitem[{Taylor et~al.(2013)Taylor, Nicolle, and Maguire}]{Taylor2013-lh}
Shena~Parthab Taylor, Colette Nicolle, and Martin Maguire. 2013.
\newblock Cross-cultural communication barriers in health care.
\newblock \emph{Nurs Stand}, 27(31):35--43.

\bibitem[{Wei et~al.(2022)Wei, Wang, Schuurmans, Bosma, Ichter, Xia, Chi, Le,
  and Zhou}]{wei2022}
Jason Wei, Xuezhi Wang, Dale Schuurmans, Maarten Bosma, Brian Ichter, Fei Xia,
  Ed~H. Chi, Quoc~V. Le, and Denny Zhou. 2022.
\newblock Chain-of-thought prompting elicits reasoning in large language
  models.
\newblock In \emph{Proceedings of the 36th International Conference on Neural
  Information Processing Systems}, NIPS '22, Red Hook, NY, USA. Curran
  Associates Inc.

\bibitem[{Xian et~al.(2024)Xian, Chang, Xiang, and Liu}]{Xian2024-go}
Xuechang Xian, Angela Chang, Yu-Tao Xiang, and Matthew~Tingchi Liu. 2024.
\newblock Debate and dilemmas regarding generative {AI} in mental health care:
  Scoping review.
\newblock \emph{Interact J Med Res}, 13:e53672.

\bibitem[{Xu et~al.(2024)Xu, Sharaf, Chen, Tan, Shen, Van~Durme, Murray, and
  Kim}]{xuetal2024}
Haoran Xu, Amr Sharaf, Yunmo Chen, Weiting Tan, Lingfeng Shen, Benjamin
  Van~Durme, Kenton Murray, and Young~Jin Kim. 2024.
\newblock Contrastive preference optimization: pushing the boundaries of llm
  performance in machine translation.
\newblock In \emph{Proceedings of the 41st International Conference on Machine
  Learning}, ICML'24. JMLR.org.

\bibitem[{Xu et~al.(2025)Xu, Cook, Semaan, and Voida}]{xuetal2025}
Tian Xu, Sid Cook, Bryan Semaan, and Stephen Voida. 2025.
\newblock \href {https://doi.org/10.1145/3706598.3713669} {Expression-in-action
  and expression-on-action: A systematic review of mediums for expression in
  mental health}.
\newblock In \emph{Proceedings of the 2025 CHI Conference on Human Factors in
  Computing Systems}, CHI '25, New York, NY, USA. Association for Computing
  Machinery.

\bibitem[{Zhang et~al.(2025)Zhang, Zhang, Zhou, Louie, Kim, Guo, Li, and
  Peng}]{zhangetal2025}
Han Zhang, Jiaqi Zhang, Yuxiang Zhou, Ryan Louie, Taewook Kim, Qingyu Guo,
  Shuailin Li, and Zhenhui Peng. 2025.
\newblock \href {https://doi.org/10.1145/3711031} {Mentalimager: Exploring
  generative images for assisting support-seekers' self-disclosure in online
  mental health communities}.
\newblock \emph{Proc. ACM Hum.-Comput. Interact.}, 9(2).

\end{thebibliography}


\end{document}